\documentstyle[prb,aps,epsfig,multicol]{revtex}
\def\high{high-$T_c$ }
\begin{document}
\draft
\title{
Quasiparticle spectrum in the vortex state of $d-$wave
superconductors}
\author{H. Won,$^{1,2}$ and K. Maki$^{2}$}
\address{$^1$Department of Physics, Hallym University,
Chunchon, 200-702, South Korea \\
$^2$Department of Physics and Astronomy,
University of Southern California,
Los Angeles, CA 90089-0484}
\date{\today}
\maketitle
\begin{abstract}
Quasiparticle spectrum associated with
the nodal structure in $d$-wave superconductors
is of great interest.
We study theoretically the quasi-particle spectrum in a planar
magnetic field, where the effect of the magnetic field is treated 
in terms of the Doppler shift.
We  obtain the angular  dependent specific heat
in the presence of a planar magnetic field and impurities,
both in
the superclean limit($\frac\Gamma\Delta
\ll \frac{H}{H_{c2}} \ll 1$)  and 
in  the clean limit($\frac{H}{H_{c2}} \ll 
\frac\Gamma\Delta \ll 1$).  Also  a
similar
analysis is used for  the thermal conductivity  tensor within the
$a$-$b$ plane. 
In particular, in contrast to the earlier
works, we
find  a fourfold symmetry term in $\kappa_{\parallel}$ and 
$\kappa_{\perp} \sim -H \sin(2\theta)$ where  
$\kappa_\parallel$ and 
$\kappa_{\perp}$ are the diagonal- and the off-diagonal components 
of the thermal
conductivity tensor and 
$\theta$ is the angle between  the heat current and
the  magnetic
field.
\end{abstract}


\begin{multicols}{2}
\narrowtext

\section{Introduction}
\label{sec1}
In the last few years $d$-wave superconductivity in hole-doped 
high-$T_c$
cuprates is  established.\cite{harlingen,tsuei1}  
More  recently,  Tsuei and  Kirtley and other groups
\cite{tsuei2}
have
shown  by means of the phase sensitive  tri-crystal geometry and
the magnetic penetration depth measurement
that
the
superconductivity in the  electron doped \high cuprates  NCCO
and
PCCO is also  of $d$-wave. This  suggests strongly the
universality of
$d$-wave superconductivity   for all  \high   cuprate
superconductors.
Further, recent   measurements of  the  magnetic penetration depth in
organic  superconductor 
$\kappa$-(BEDT-TTF)$_2$Cu(N(CN)$_2$)Br  
indicates  this  system  has  also
$d$-wave
superconductivity, \cite{carrington,pinteric} 
though   the phase  sensitive  experiment
is   not
available for this system. 
Also uncertain is the nodal directions of
the order parameter \cite{schrama,ichimura} 
in $\kappa$-(ET)$_2$ salts in spite  of 
the  theoretical suggestion that it
belongs to $d_{xy}$-symmetry. \cite{visentini,schmalian,louati}
In this circumstance, the experimental probe  which indicates the
nodal
directions in $d$-wave superconductors is of prime importance. 
\par
We recall
earlier  study of  the  thermal conductivity   tensor
in  the
optimally doped YBCO in a planar  magnetic field indicates
clearly the
nodal   directions   of   $d$-wave   superconductors   consistent
with
$d_{x^2-y^2}$ symmetry. \cite{salamon,yu,aubin}  

In the meanwhile Volovik\cite{volovik} made the crucial
observation that the thermodynamics of the 
vortex state in $d$-wave superconductors is dominated by the
extended quasi-particle states associated with the
nodal structure and the effect of the magnetic field is
incorporated by the Doppler shift in the energy
spectrum.\cite{maki1}
This approach is further developed by
Barash et al\cite{barash} 
and K\"{u}bert et al.\cite{kubert1,kubert2}

In particular, those authors  consider the supercurrent
associated with individual vortices explicitly and then made
spatial average of individual vortex contribution over the unit
cell of the vortex lattice.
Indeed, 
K\"{u}bert et al have succeeded not only to
describe the
early specific heat  data by K.  Moler et  al
and B. Revaz et al  \cite{moler,revas-cz}, 
but also
predicted the
thermal conductivity in a  magnetic field parallel  to the 
$c$-axis
in the
low temperature limit,  which is  confirmed  by the
thermal
conductivity measurement by Chiao et al. \cite{chiao} 
More recently, this
method
has been used to calculate  a variety of quantities,  which
includes the
angular dependence of the specific heat of $d$-wave
superconductivity in
a planar magnetic field. \cite{vekhter1,vekhter2}

The object of this paper is to study both  the thermodynamics and
the
thermal conductivity tensor in  a planar magnetic  field. 
We find  the analytic expressions  of the density  of
states
and the thermal conductivity tensor in the presence of a magnetic
field and impurity.
Also, the Doppler shift is handled consistently with the layered
structure
common to both  
\high cuprates  and  $\kappa$-(ET)$_2$  salts.\cite{won1} 
This  enables us
to
analyze these quantities 
both in
the superclean limit ($\frac\Gamma\Delta
\ll \frac{H}{H_{c2}} \ll 1$)  and in  the clean limit
($\frac{H}{H_{c2}} \ll \frac\Gamma\Delta \ll 1$), while
in [16-18] they are called the clean limit and the dirty limit.
We think the word "dirty limit" should  not  be used to the system where
$\frac\Gamma\Delta \ll 1$.
Here $H_{c2}$ is the upper critical field, $\Gamma$ the quasi-particle
scattering rate, and $\Delta$ the order parameter.
We find the angular dependence of both the
specific
heat and the superfluid density  roughly by a factor 4  less than
those
predicted in [23].  
Also, the longitudinal thermal  conductivity
increases  linearly with $H$ in the superclean limit, 
while $H\ln(H_{c2}/H)$ in the clean limit.
Therefore, our result is similar to [18] in the clean limit, 
though it is different
from [18] in the superclean limit. 
So the thermal conductivity increases with $H$ as
in [18] but in contradiction to [13].
Further  the thermal conductivity
exhibits a
similar $\theta$-dependence as observed in [13],
but again it has the opposite sign; 
$\kappa_{\parallel}$ takes the minimum value for $\theta=\pm \frac\pi4$
contrary to [13].
Perhaps this is due to the fact there is more quasi-particle for
$\theta=0$ than for $\theta=\pm \frac\pi4$.
On the other hand, the transverse thermal conductivity is
described by
$\kappa_{\perp} \sim -H \sin(2\theta)$ in the both limits.
This  dependence
appears
to describe  quite well  
the data reported in [11,12]. 
For simplicity, we
limit our
analysis to 
$T \ll T_c$ and $H\ll H_{c2}$ and the impurity scattering is treated in the
unitarity
limit. \cite{hotta,sun}
\section{Density of States, specific heat, and superfluid density}
\label{se2}
Following Barash et al,\cite{barash} the impurity-renormalized
quasiparticle energy $\tilde{\omega}$
is given by
\begin{equation}
\tilde{\omega} =  \omega + i \frac{\Gamma}{g(\omega)} 
\end{equation}
and
\begin{equation}
g(\omega) =  \langle  \frac{\tilde{\omega}- {\bf v}\cdot {\bf q}}
{\sqrt{(\tilde{\omega} -{\bf v}\cdot {\bf q})^2 - \Delta^2
\cos^2(2\phi) }}\rangle 
\end{equation}
where 
$|{\bf v}\cdot{\bf q}|$
=$\displaystyle \frac{1}{2r}$ 
$\sqrt{v v^{\prime} (\sin^2\chi +\sin^2(\phi-\theta))} $,
the Doppler-shifted energy due to the circulating supercurrent
around the vortex, $r$ the distance from the vortex,
$v$ and $v^{\prime}$ are the Fermi velocity  within the
$a$-$b$ plane and the one parallel to the $c$-axis, respectively,
where the effect of the layered structure is explicitly
considered.\cite{won1} 
Here $\chi=cp_3$,
$ \phi$ is the angle the quasi-particle
momentum in the $a$-$b$ plane makes from the $a$-axis,
$\theta$ is the angle the magnetic field makes from the $a$-axis,
and $< \ldots> = \frac{1}{(2\pi)^2}
\int d\chi \int d \phi \ldots$ the angular average.
We assume the unitarity limit of the impurity scattering.
The low temperature  limit of both  the specific  heat and the
superfluid
density is easily obtained from the residual density of states
\cite{won2}
(i.e. the
density of states on the Fermi surface or at $\omega=0$).
The quasi-particle density of
states at $\omega=0$ in the presence of both impurities and a magnetic
field is given by
\begin{eqnarray}
\frac{N(\omega=0)}{N_0} &=& {\rm Re}\,\,\, g(\omega=0)  
   \nonumber \\
 & = & \frac2\pi
\Big\langle  C_0 \ln (\frac{4}{\sqrt{C_0^2 +x^2}})
+ x\tan^{-1} (\frac{x}{C_0} )\Big\rangle 
\end{eqnarray}
where $x=|{\bf v} \cdot {\bf q}| /\Delta$ and  we used
\begin{equation}
\frac{\tilde{\omega}}{\Delta}\big|_{\omega=0} = i C_0 =
\frac{i\Gamma}{\Delta g(\omega=0)}
\end{equation}
This set of equation is solved in the following: 

\par
For $\quad C_0 \ll \langle x \rangle \ll 1$ 
(i.e.\,\,\,$\frac\Gamma\Delta
\ll \frac{H}{H_{c2}} \ll 1$),  
\begin{eqnarray}
\frac{N(\omega=0)}{N_0}& & \!\!\!\big(
\equiv\frac{N(H, \theta)}{N_0}\,\big)
\nonumber\\
&= &\langle x \rangle + \frac{2}{\pi} 
\frac{\Gamma}{\Delta }\frac{1}{\langle x\rangle }
\langle\ln (\frac4x) -1 \rangle , 
\end{eqnarray}
and
\[
 C_0 =\frac\Gamma\Delta \frac1{\langle x \rangle} + \ldots
\]
\par
For $\langle x \rangle  \ll C_0  \ll 1 $
(i.e.\,\,\, $\frac{H}{H_{c2}} \ll \frac\Gamma\Delta \ll 1$),
\begin{equation}
 \frac{N(\omega=0)}{N_0}\big(\equiv\frac{N(H, \theta)}{N_0}\,\big)
  = \frac{N_{\rm imp}(0) }{N_0}
\big( 1 + \frac{1}{2} \frac{\Delta}{\Gamma} \langle  x^2\rangle \big)
\end{equation}
and
\[
C_0 \simeq \sqrt{\frac\pi2\frac\Gamma\Delta/
\ln(4\sqrt{\frac2\pi \frac\Delta\Gamma})}
\]
where $N_{\rm imp}(0) $  is the density of states in the $H=0$ case
with  the unitarity  impurity scatterers and is given \cite{sun}
\begin{equation}
\frac{N_{\rm imp}(0) }{N_0}
=\frac{2}{\pi} C_0 \ln (\frac4{C_0})
\simeq \sqrt{\frac2\pi \frac\Delta\Gamma \ln(4\sqrt{
\frac2\pi \frac\Gamma\Delta })}
\end{equation}
We call the former the superclean limit while the latter the
clean limit.
These have been obtained essentially in [16] except for a few
typos. We used  the same 
angular and the spatial average
as in   Barash et al and K\"{u}bert et al.
\cite{barash,kubert1}
Finally, the density of states in the both limits
is the following:
\par
For the superclean limit ($C_0 \ll \langle x\rangle  \ll 1$);
\begin{eqnarray}
\frac{N(H,\theta)}{N_0} 
&\simeq&
\frac{\sqrt{vv^{\prime} e H}}{\Delta} I(\theta)
+
\frac{2}{\pi} 
\frac{\Gamma}{\sqrt{vv^{\prime} e H}}\frac{1}{I(\theta)}
\big[
\ln (\frac{8\Delta}{\sqrt{vv^{\prime} e H}}) 
\nonumber \\
&-&\frac32 -J(\theta)]
\end{eqnarray}

where 
\begin{eqnarray}
I(\theta) &=& \frac12 \Sigma_{\pm} \Big\langle 
\sqrt{\sin^2\chi + \sin^2(\pm
\frac\pi4 -\theta)}\,\,\Big\rangle 
\nonumber \\
&=&
\frac1\pi \Sigma_{\pm}^{\ } \sqrt{\frac{3\pm s}{2}}
E(\sqrt{\frac{2}{3\pm s}})
\nonumber\\
&\simeq& 0.0285\,\cos(4\theta) + 0.955
\end{eqnarray}
and 
\begin{eqnarray}
J(\theta) &=&\frac12\Sigma_{\pm}
\Big\langle \ln \sqrt{\sin^2\chi + \sin^2(\pm\frac\pi4 -\theta)}
\,\,\Big\rangle
\nonumber \\
&=&\frac14 \Sigma_{\pm} \ln
\frac{2\pm s +
\sqrt{(3\pm s)(1\pm s)}}{4}
\nonumber\\
&\simeq& -0.778
+0.744\,{\rm Max}\{|\cos\theta|, \,\,|\sin\theta|\}
\end{eqnarray}
where $s$ = $\sin(2\theta)$
and $E(k)$ is the complete elliptic integral. 
Here, instead of average over $\phi$, we average over
$\phi=\pm\frac\pi4$
(i.e. over the two nodal directions. \cite{vekhter2})
\par
The $\theta$-dependence of $I(\theta)$ and $J(\theta)$ are shown
in Fig.1 and Fig.2.
In Fig.1 we compared also the angular dependent obtained in [23]
where the layered structure is ignored.
In the present calculation, the angular dependent term is 
 about the 4$\%$ of the coefficient $\sim \sqrt{H}$, while in [23] it is about 
20 $\%$.
A recent specific data from 
stoichiometric YBa$_2$Cu$_3$O$_{7.00}$ crystals \cite{revas}
appear to be more consistent with the present analysis.
Perhaps we have to point out the angular dependence of $J(\theta)$ 
appeared since we cut off the
the short-range logarithmic divergence at 
$x=\frac12$ where the uniform order parameter started to
be greatly disturbed,\cite{won3}
though the exact value $\frac12$ in the present case 
is not so important.
\par
For the clean limit ($\langle x \rangle\ll C_0 \ll 1$);
\begin{equation}
\frac{N(H,\theta)}{N_0}
 = \frac{N_{\rm imp}(0)}{N_0}
[1 + \frac{1}{4} \frac{v v^{\prime} e H}{\Gamma\Delta}
( \ln (\sqrt{\frac{\Delta}{v v^{\prime} eH}}) 
-F(\theta) ) \big]
\end{equation}
where
\begin{eqnarray}
F(\theta) &=& 
\frac12 \Sigma_{\pm}
\Big\langle (\sin^2\chi +\sin^2(\pm \frac\pi4 -\theta)) \times
\nonumber \\
&  &\ln\sqrt{\sin^2\chi +\sin^2(\pm \frac\pi4 -\theta)}\,\,\, \Big\rangle
\nonumber\\
&=& \frac14 
\Sigma_{\pm} \Big[
(2\pm s) \ln
(\frac{2 \pm  s +
\sqrt{(3\pm s)(1\pm s)}}{4})
\nonumber\\
&+&4 -\sqrt{(3\pm s)(1\pm s)} \,\,\,\Big]
\nonumber \\
& \simeq&
0.147 - 0.082\,\cos(4\theta)
\end{eqnarray}
The function $F(\theta)$ is shown in Fig.3. 
\par
The residual density of states is most readily accessible to the
low temperature limit of the spin
susceptibility as in NMR, the specific heat and the superfluid
density.\cite{won2}
In particular the low temperature specific heat and the 
superfluid density are given by
\begin{equation}
C_s(H, \theta) =
\frac{2 \pi^2}{3} TN(H, \theta)
\end{equation}
and
\begin{equation}
\rho_s(H,\theta)= 1 -N(H,\theta)/N_0
\end{equation}
Also, in the superclean limit and $T/\Delta > \langle x \rangle$, we obtain
\cite{won4}
\begin{equation}
C_s = 18\zeta(3) \frac{T^2}\Delta N_0
\end{equation}
and
\begin{equation}
\rho_s(T)/\rho(0) = 1 -2(\ln2)\frac T\Delta
\end{equation}
To conclude this section, we have considered the residual density of states 
in a planar magnetic field both in the superclean and 
in the clean limit.
In both limits the residual density of states 
exhibits the fourfold symmetry,
though magnitude of this terms is roughly 4 times smaller 
than the earlier
result.\cite{vekhter2}

\section{Thermal Conductivity tensor}
Making use of Ambegaokar-Griffin formula \cite{ambegaokar}  
the  low temperature thermal
conductivity tensor  
is given by

\begin{equation}
\kappa_{\parallel}/\kappa_{0} 
 =  
\frac\pi2 \Big\langle  \frac{{\displaystyle\frac12}
\Big(1 + \frac{\displaystyle C_0^2 + x^2 -\cos^2(2\phi)}
{\displaystyle |(C_0 + ix)^2
+\cos^2(2\phi)|} \Big)} 
{{\rm Re}\sqrt{(C_0 + i x)^2 +\cos^2(2\phi)}} \Big \rangle 
\end{equation}
Similarly
\begin{equation}
\kappa_{\perp}/\kappa_0
 =   
\frac\pi2 \Big\langle  \sin(2\phi) \frac{{\displaystyle\frac12}
\Big(1 + \frac{\displaystyle C_0^2 + x^2 -\cos^2(2\phi)}
{\displaystyle |(C_0 + ix)^2
+\cos^2(2\phi)|}\Big)} 
{{\rm Re}\sqrt{(C_0 + i x)^2 +\cos^2(2\phi)}} \Big \rangle
\end{equation}
where $\kappa_0 =\kappa_{\parallel}(H=0)$
\par
First, for the superclean limit ($C_0 \ll \langle x 
\rangle \ll 1$)
Eq.(17) and Eq.(18) reduce to 
\begin{eqnarray}
\kappa_{\parallel}/\kappa_n &\simeq& \frac2\pi \langle x\rangle^2
\nonumber\\
&=& \frac2\pi \frac{v v^{\prime} eH}{\Delta^2} \big(I(\theta)\big)^2
\end{eqnarray}
and 
\begin{equation}
\kappa_{\perp}/\kappa_n = -\frac2\pi 
\frac{v v^{\prime} eH}{\Delta^2} I(\theta)L(\theta)
\end{equation}
where 
$\kappa_n=\displaystyle 
\frac{\pi^2 T n}{6\Gamma m}$, the thermal conductivity in the
normal state, and 
\begin{eqnarray}
L(\theta) &=& \frac1\pi \big(\sqrt{\frac{3+s}{2}}E(\sqrt{\frac{3+s}{2}})
-\sqrt{\frac{3+s}{2}}E(\sqrt{\frac{3+s}{2}}) \big) 
\nonumber \\
&\simeq & 0.29\,\,\sin(2\theta)
\end{eqnarray}
The function $L(\theta)$ is shown in Fig.4 together with the approximate
form.
We note also that $\kappa_{\parallel}$ has the same 
angular dependence as $[N(H,\theta)]^2$.
Also $\kappa_{\perp}$ is proportional to $\sin(2\theta)$ 
in a good approximation (see Fig.4 and Eq.(9)).
\par
On the other hand, in the clean limit($\langle x\rangle  \ll C_0 \ll 1$),
we obtain
\begin{eqnarray}
\kappa_{\parallel}/\kappa_0 &=& 1 + \frac13 \frac{\langle
x^2 \rangle}{C_0^2} \nonumber \\
&=& 1+ \frac1{3\pi}
\frac{v v^{\prime} eH}{\Gamma \Delta}
\sqrt{\ln(4\sqrt{\frac{2\Delta}{\pi\Gamma}})}
\Big[ \ln
(\frac{2\Delta}{\sqrt{v v^{\prime} e H}}) 
\nonumber \\
&-& F(\theta)
\,\,\Big]
\end{eqnarray}
 where $F(\theta)$ is given in Eq.(12).
Note that $F(\theta)$ appears from the
lower cut-off of $r$ as explained after
Eq.(10).
And \begin{eqnarray}
\kappa_{\perp}/\kappa_0 &=&
-\frac1{3\pi}
\frac{v v^{\prime} eH}{\Gamma \Delta}
\sqrt{\ln(4\sqrt{\frac{2\Delta}{\pi\Gamma}})}
\Big[ \sin(2\theta) \ln 
(\frac{2\Delta}{\sqrt{v v^{\prime} e H}}) 
\nonumber \\
&-& G(\theta)
\,\,\Big]
\end{eqnarray}
where
\begin{eqnarray}
G(\theta) &=&
\frac14
 \Big[
(2+ s) \ln
(\frac{2 +  s +
\sqrt{(3+ s)(1 + s)}}{4}) 
\nonumber \\
&- &(2- s) \ln
(\frac{2 -  s +
\sqrt{(3- s)(1 - s)}}{4})
+ 2\sin(2\theta)
\nonumber\\
& -&\sqrt{(3+ s)(1+ s)} 
+\sqrt{(3- s)(1- s)} \,\,\,\Big]
\nonumber\\
&\simeq& 0.422 \sin(2\theta)
\end{eqnarray}
\noindent
The function $G(\theta)$ is shown in Fig.5 with the approximate
form.
Here, $\kappa_0=\displaystyle \frac\pi3 \frac{T n}{\Delta m}$,
$n$ is  the quasiparticle density and $m$ the quasiparticle mass. 
This form of thermal conductivity, in the absence of magnetic field is
derived
first by Lee. \cite{lee}
In general, however, $\Delta$ in $\kappa_0$ depends on both on
$\Gamma$ and $H$.
In the superclean limit, $\Delta(H)$ may be approximately given by
\cite{won3}
\begin{eqnarray}
\Delta(H)/\Delta_0 &=& 1-\frac13
\langle x^3 \rangle,
\quad {\rm for}\quad  C_0 \ll \langle x \rangle  \ll 1 
\nonumber \\
&\simeq& 1 -\frac16 \frac{v v^{\prime} e H}{\Delta^2}
\end{eqnarray}
In the clean limit ($ \langle x \rangle  \ll C_0 \ll 1$),
\begin{equation}
\Delta(\Gamma)/\Delta_0 \simeq 1-\frac\pi4 \frac\Gamma\Delta
\end{equation}
\par 
In both limits, the longitudinal conductivity 
increases with $H$, linearly in superclean limit 
and $H\ln (H_{c2}/H)$ in clean limit, respectively.
(Our result agrees with [18] only in the clean limit, 
$\frac{H}{H_{c2}} \ll \frac\Gamma\Delta \ll 1$.)
Further it exhibits 
the $\theta$ dependence in both
the superclean and the clean limit.
In the superclean limit the $\theta$-dependence 
comes from that of the density of states,
while in the clean limit this arises from the short range cut-off
we have introduced after Eq.(10).   
Also this angular dependence 
is very similar to 
the one
reported in [13], but of opposite sign.
But perhaps of particular interest is the transverse thermal conductivity.
The dominant terms  in Eq.(20) and Eq.(23),
exhibit $\sin(2\theta)$ dependence,
which is fully consistent with the early experiment.
\cite{salamon,yu}
Also this $\sin(2\theta)$ dependence is appreciable
even when $H \sim H_{c2}$.\cite{maki2}

\section{Concluding Remarks}
We have  extended earlier  analysis  of the  thermodynamics and
the
transport properties in  d-wave superconductors in  2
directions.  First,
we  take   account  of   the  layered  structure   of  the
underlying
superconductors  explicitly.   Second,  we   focused on   the
angular
dependence of thermal conductivity  tensor, which exhibits
clear signs
of the nodal  structures in d-wave  superconductors. 
Indeed the diagonal thermal conductivity exhibits the
fourfold symmetry as observed in [13], but of opposite 
sign.
In
particular,  the
present  result   describes the  transverse thermal
conductivity  (or
Righi-Leduc effect)  observed in  YBCO. \cite{yu,aubin} 
Indeed,  we have
found
recently   a  similar   transverse  thermal   conductivity  in
p-wave
superconductors as Sr$_2$RuO$_4$  and   
$f$-wave superconductors as  in
UPt$_3$.\cite{maki2}
Therefore, this $\sin(2\theta)$  dependence in $\kappa_{\perp}$ is
rather  common to
most of
unconventional superconductors.

In summary, the  exploration of the  nodal structure in
unconventional
superconductors  will  provide  useful   insight in   the
quasi-particle
spectrum in the vortex state.

\acknowledgements
We are benefited from  discussions with P. Esquinazi,  T.
Ishiguro, K.
Izawa, Y. Maeno, Y. Matsuda, B. Revas and M.A. Tanatar on
ongoing
experiment on \high  cuprates and on 
Sr$_2$RuO$_4$. 
HW acknowledges the support from the 
Korean Science and Engineering Foundation (KOSEF)
through the Grant No. 1999-2-114-005-5.
Also, HW thanks Dept. of Physics and Astronomy, USC
for their hospitality during her stay.

\begin{figure}
\caption{$I(\theta)$ and the approximate form are shown as
function of $\theta$. $I_1(\theta)$ is the angular dependence in
[23] where the layer structure is ignored.}
\end{figure}

\begin{figure}
\caption{$J(\theta)$ is shown as a function of $\theta$.
It has cusps at $\pi/4$ and $3\pi/4$ together with 
the approximate form. }
\end{figure}

\begin{figure}
\caption{$F(\theta)$ is shown as a function of $\theta$
together with the approximate form.
This angular dependence appears in the specific heat
$C_v$, the 
superfluid density $\rho_s$, and thermal conductivity tensor
$\kappa_{\parallel}$ in clean limit($\frac{H}{H_{c2}} \ll
\frac\Gamma\Delta \ll 1$).}
\end{figure}
 
\begin{figure}
\caption{$L(\theta)$ is shown as a function 
of $\theta$ to together with the approximate form.}
\end{figure}

\begin{figure}
\caption{$G(\theta)$ is shown as a function of $\theta$
together with the approximate form.}
\end{figure}

\end{multicols} 

\end{document}